\documentclass[iop]{emulateapj}

\slugcomment{}

\shorttitle{GRBs from BH-BH mergers}
\shortauthors{Zhang}

\begin{document}


\title{Mergers of Charged Black Holes: Gravitational Wave Events, Short Gamma-Ray Bursts, and Fast Radio Bursts}


\author{Bing Zhang}
\affil{Department of Physics and Astronomy, University of Nevada Las Vegas, NV 89154, USA}



\begin{abstract}
The discoveries of GW 150914, GW 151226, and LVT 151012 suggest that double black hole (BH-BH) mergers are common in the universe. If at least one of the two merging black holes carries certain amount of charge, possibly retained by a rotating magnetosphere, the inspiral of a BH-BH system would drive a global magnetic dipole normal to the orbital plane. The rapidly evolving magnetic moment during the merging process would drive a Poynting flux with an increasing wind power. The magnetospheric activities during the final phase of the merger would make a fast radio burst (FRB) if the BH charge can be as large as a factor of $\hat q \sim (10^{-9}-10^{-8})$ of the critical charge $Q_c$ of the BH. At large radii, dissipation of the Poynting flux energy in the outflow would power a short duration high-energy transient, which would appear as a detectable short-duration gamma-ray burst (GRB) if the charge can be as large as $\hat q \sim (10^{-5}-10^{-4})$. The putative short GRB coincident with GW 150914 recorded by Fermi GBM may be interpreted with this model. Future joint GW/GRB/FRB searches would lead to a measurement or place a constraint on the charges carried by isolate black holes. 
\end{abstract}





\section{Introduction}

Black holes (BHs) are uniquely described with three parameters, mass $M$, angular momentum $J$, and charge $Q$. Whereas the first two parameters have been measured with various observations for both stellar-mass and super-massive BHs, it has been widely believed that the $Q$ parameter must be very small. However, no measured value or upper limit of $Q$ have been reported for any BH.

Recently, the Laser Interferometer Gravitational-wave Observatory (LIGO) team announced the ground-breaking discovery of the first gravitational wave (GW) source, GW 150914, which is a BH-BH merger with two BH masses $36^{+5}_{-4} M_\odot$ and $29^{+4}_{-4} M_\odot$, respectively \citep{GW150914}. Two other BH-BH merger events (GW 151226 and LVT 151012) were later announced \citep{GW151226}.  The inferred event rate density of BH-BH mergers is $\sim (9-240) ~{\rm Gpc}^{-3}~{\rm yr}^{-1}$ \citep{GW-rate}. Intriguingly, the {\em Fermi} GBM team reported a 1-second long, putative weak gamma-ray burst (GRB) 0.4 seconds after the GW event was detected (\cite{connaughton16}, but see \cite{greiner16,xiong16}). This is surprising, since unlike NS-NS and NS-BH mergers which can form BH-torus systems and produce short GRBs through accretion \citep{paczynski86,eichler89,paczynski91,narayan92,meszarosrees92,rezzolla11}, BH-BH mergers are not expected to have enough surrounding materials with a high enough density to power a short-duration GRB via accretion.

On the other hand, fast radio bursts (FRBs) are mysterious milliseconds-duration radio transients \citep{lorimer07,thornton13}. Recent observations suggest that at least some FRBs are likely at cosmological distances \citep[e.g.][]{keane16}. Their physical origins, however, remain unknown.

Here we show that if at least one BH in the two merging BHs carries a certain amount of charge, the inspiral of the BH-BH system would induce a global magnetic dipole normal to the orbital plane. The rapid evolution of the magnetic moment would drive a Poynting flux with an increasing wind power, which may give rise to an FRB and even a  short-duration GRB depending on the value of the charge. 

\section{Electrodynamics of charged black hole merger system}

For a charged black hole, one can define the Schwarzschild radius and the Reissner-Nordstr\"om (RN) radius
\begin{equation}
r_s = \frac{2GM}{c^2}, ~~ r_Q = \frac{\sqrt{G} Q}{c^2}, 
\end{equation}
where $M$, $Q$ are the mass and charge of the black hole, respectively, $G$ and $c$ are the gravitational constant and speed of light, respectively, and the electrostatic cgs units have been used. By equating $r_s$ and $r_Q$, one may define a characteristic charge
\begin{equation}
Q_c \equiv 2 \sqrt{G} M = (1.0 \times 10^{31}~{\rm e.s.u.}) \left(\frac{M}{10 M_\odot}\right),
\label{eq:Qc}
\end{equation}
which is $(3.3 \times 10^{21}~{\rm C})~ (M/10 M_\odot)$ in the S.I. units. The charge of this magnitude would significantly modify the space-time geometry with a magnitude similar to $M$. We consider a BH with charge 
\begin{equation}
Q = \hat q Q_c, 
\end{equation}
with the dimensionless parameter $\hat q \ll 1$. For simplicity, in the following we consider two identical BHs with the same $M$ and $Q$.

As the two BHs spiral in\footnote{For an order-of-magnitude treatment, we apply classical mechanics and electrodynamics without general relativity correction.}, a circular current loop forms, which gives a time-dependent magnetic dipole moment
\begin{eqnarray}
\mu & = & \frac{\pi I (a/2)^2}{c} =  \frac{\sqrt{2 G M a} Q}{4c} = \frac{\sqrt{2} G^{3/2} M^2}{c^2} \hat q {\hat a}^{1/2} \nonumber \\
& = & (1.1 \times 10^{33}~{\rm G~cm^3})  \left(\frac{M}{10 M_\odot}\right)^2 \hat q_{-4} \hat a^{1/2},
\label{eq:mu}
\end{eqnarray}
where $I = 2Q/P$ is the current, and
\begin{eqnarray}
P & = & \frac{2\pi}{\sqrt{2GM}} a^{3/2} = 8\sqrt{2}\pi \frac{GM}{c^3} \hat a^{3/2} \nonumber \\
& = & (1.7~{\rm ms}) \left(\frac{M}{10M_\odot}\right) \hat a^{3/2}
\end{eqnarray} 
is the Keplerian orbital period, $a = \hat a (2 r_s)$ is the separation between the two BHs, and $\hat a$ is the distance normalized to $2 r_s$. Notice that at the coalescence of the two BHs, $\hat a=1$ for two Schwarzschild BHs, but $\hat a$ can be as small as 0.5 for extreme Kerr BHs. For comparison, a magnetar with a surface magnetic field $B_p \sim 10^{15}$ G and radius $R_{\rm NS} \sim 10^6$ cm has a magnetic dipole $\mu_{\rm mag} \sim B_p R_{\rm NS}^3  = (10^{33} ~{\rm G~cm^3}) B_{p,15} R_{\rm NS,6}^3$.

The orbital decay rate due to gravitational wave radiation can be generally written as $da/dt = - (64/5) G^3 {\cal M} M_{\rm tot}^2/ [c^5 a^3 (1-e^2)^{7/2}] (1+(73/24) e^2 + (37/96)e^4$, where ${\cal M} = M_1 M_2/ M_{\rm tot}$ is the chirp mass, and $M_{\rm tot} = M_1 + M_2$ is the total mass of the system. Assuming $M_1=M_2$ for simplicity and adopting $e=0$ which is valid before the coalescence, one gets
\begin{equation}
 \frac{da}{dt} = -\frac{2}{5} \frac{c}{\hat a^3}.
\end{equation}
The rapid evolution of the orbital separation before the coalescence leads to a rapid change of the magnetic flux, and hence, a Poynting flux with increasing power. A full description of the electrodynamics of the system requires numerically solving Einstein equations with electrodynamics. To an order of magnitude analysis, one may estimate the Poynting flux wind luminosity using a magnetic dipole radiation formula in vacuum, i.e.
\begin{eqnarray}
 L_w & \simeq &  \frac{2 \ddot\mu^2}{3 c^3} \simeq \frac{49}{120000} \frac{c^5}{G} \hat q^2 \hat a^{-15} \nonumber \\
 & \simeq & (1.5\times 10^{48}~{\rm erg~s^{-1}}) \hat q_{-4}^2 \hat a^{-15},
 \label{eq:Lw1}
\end{eqnarray}
where $\ddot \mu$ is the second derivative of the magnetic dipole moment $\mu$.
Notice that this wind power {\em is determined by fundamental constants and the dimensionless parameters $\hat q$ and $\hat a$ only}. 
Noticing that the gravitational wave radiation power can be estimated as
\begin{eqnarray}
L_{\rm GW} & \simeq & \frac{c^5}{G} \left(\frac{GM}{c^2 a}\right)^5 = \frac{1}{1024}\frac{c^5}{G} \hat a^{-5}, \nonumber \\
& \simeq & (3.6\times 10^{56}~{\rm erg~s^{-1}}) \hat a^{-5},
\end{eqnarray}
one can also write
\begin{equation}
L_w \sim 0.4 \hat q^2 L_{\rm GW} \hat a^{-10}.
\end{equation}

One may show that particles can be accelerated to a relativistic speed from the global magnetosphere. The rapid evolution of the orbital separation before the coalescence leads to a rapid change of the magnetic flux, and hence, induce a huge electromotive force (EMF). At a relatively large distance $r$ from the merging system ($r \gg a$), one may approximate the instantaneous magnetic field configuration as $B_r = (\mu/r^3) (2 \cos\theta)$ and $B_\theta = (\mu/r^3) \sin\theta$ with the dipole moment $\mu$ expressed in Eq.(\ref{eq:mu}). The magnetic flux through the upper hemisphere with radius $r$ is $\Phi = \int_0^{\pi/2} 2\pi r^2 \sin\theta (\mu/r^3)(2\cos\theta) d\theta = 2\pi \mu/r$.  Faraday's law of magnetic induction then gives an induced EMF
\begin{eqnarray}
 {\cal E}  & = & -\frac{1}{c}\frac{d\Phi}{dt} = -\frac{2\pi}{cr} \frac{d\mu}{dt} = \frac{\sqrt{2}\pi}{10} \frac{G^{1/2}M}{r}\hat q \hat a^{-7/2} 
\label{eq:emf}
\end{eqnarray}

Similar to the case of a rotation-powered pulsar, such an EMF across different field lines would lead to particle acceleration and a photon-pair cascade \citep[e.g.][]{ruderman75b,arons79,muslimov92,harding98,zhangharding00}. The physical processes involved are complicated and deserve further studies. For an order-of-magnitude analysis, one may estimate the Poynting-flux wind power $L_w \sim {\cal E}^2/{\cal R}$, where ${\cal R}$ is the resistance of the magnetosphere, which may be taken as $c^{-1}$ for a conductive magnetosphere. This gives 
\begin{eqnarray}
 L_w & \sim & {\cal E}^2 c = \frac{\pi^2}{50} \frac{GM^2}{r^2} c \hat q^2 \hat a^{-7}
   \simeq  \frac{\pi^2}{200} \frac{c^5}{G} \hat q^2 \hat r^{-2} \hat a^{-7},
 \label{eq:Lw2}
\end{eqnarray}
where $\hat r = r/2 r_s$ is the normalized wind-launching radius. Notice that Eq.(\ref{eq:Lw2}) has the same scaling $\propto (c^5/G) \hat q^2$ as Eq.(\ref{eq:Lw1}), even though the dependence on $\hat a$ may be different (pending on how $\hat r$ depends on $\hat a$). In the following, for simplicity, we apply the vacuum formula Eq.(\ref{eq:Lw1}) to perform related estimates.

The wind power is very sensitive to $\hat a$, and increases rapidly as the orbital separation shrinks. The highest power happens right before the final merger, so that such a merger system is a plausible engine for a {\em fast radio burst} and possibly a {\em short-duration $\gamma$-ray burst}\footnote{After the submission of this paper, \cite{liu16} proposed an alternative mechanism to produce FRBs from BH-BH merger systems through triggering an instability in the Kerr-Newman BH magnetospheres.}.

One may estimate the time scale for the orbital separation to shrink from $\hat a=1.5$ to $\hat a = 1$, during which $L_w$ increases by a factor of $\sim 440$. This is
\begin{equation}
\tau_{1.5} \lesssim 
\frac{P}{|\dot P|} = \frac{20}{3} \frac{GM}{c^3} \hat a^4 \simeq (1.7~{\rm ms})~\left(\frac{M}{10M_\odot}\right) \left(\frac{\hat a}{1.5}\right)^4,
\label{eq:tr1}
\end{equation}
where $\dot P \simeq -(192 \pi/5 c^5) (2\pi G/P)^{5/3} {M}^2 M_{\rm tot}^{-1/3} = (6\sqrt{2}\pi/5) \hat a^{-5/2}$ is the orbital decay rate for GW radiation \citep{taylor89}.

It would be informative to compare the Poynting flux power proposed in this paper (Eq.(\ref{eq:Lw1})) with some other Poynting flux powers proposed in the literature. Two relevant ones are the general-relativity-induced Poynting flux power when a BH moves in a constant magnetic field $B_0$ \citep{lyutikov11a}\footnote{In a dynamically evolving system, the assumption of constant $B_0$ is no longer valid, so that more detailed modeling is needed to perform a more accurate comparison between this power and $L_w$.} and a Poynting flux power due to the interaction between the magnetospheres of two BHs \citep{lyutikov11b}\footnote{This power does not exist if only one BH carries a magnetosphere.}. Expressing Eqs.(1) and (4) in \cite{lyutikov11b} in terms of $\hat q$ using Eq.(\ref{eq:Q}) below, we find that these two powers are both of the order of $\sim (R_{\rm lc,*}/a)^2 \hat a^{15} L_w$, where $R_{\rm lc,*} = c/\Omega_*$ is the light cylinder radius of the BHs. Noticing the strong dependence on $\hat a$. These powers are negligibly small compared with $L_w$ when $\hat a$ becomes smaller than unity.

\section{On the charge of BHs}

It is well known that a rotating point magnetic dipole carries a net charge \citep{cohen75,michel82}. In the physical model of pulsars, the difficulty was not how to make a charged neutron star, but rather how to designate a return current to make a neutron star neutral (which is not necessary in pulsar emission models) \citep{michel82}. We assume that the charged BHs in our model each possesses a magnetosphere with a dipole configuration. The magnetosphere may be attained in the not-too-distant past when the BH went through a magnetically arrested accretion phase \citep[e.g.][]{tchekhovskoy11}, and the BH is still undergoing slow ``balding'' \citep{lyutikov11}. Alternatively, the magnetosphere may be maintained by a debris disk that is circulating the BH at the time of coalescence \citep[e.g.][]{perna16,li16}. The charge maintained by an astrophysical rotating dipolar magnetosphere is approximately
\begin{equation}
Q \sim \frac{ \Omega_* \mu_*}{3c},
\label{eq:Q}
\end{equation}
where $\mu_*$ (to be differentiated from $\mu$ in Eq.(\ref{eq:mu})) is the magnetic moment of the BH dipole, and $\Omega_*$ is the angular velocity of the BH magnetosphere. This may be derived 
according to the Gauss's law for a point dipole (p. 24 of \citealt{michel82}), or through a volume integration of a Goldreich-Julian magnetosphere. 

According to Eqs. (\ref{eq:Qc}) and (\ref{eq:Q}), the rotating magnetic point dipole of individual BHs with dimensionless charge $\hat q$ should satisfy
\begin{equation}
 \mu_* \Omega_* \sim (9 \times 10^{36} ~{\rm G~cm^3~s^{-1}}) \left(\frac{M}{10 M_\odot}\right) \hat q_{-5}.
 \label{eq:muOmega}
\end{equation}
For comparison, a millisecond pulsar has $\mu_* \Omega_* \sim 10^{37}~{\rm G~cm^3~s^{-1}}$.

The spin-down luminosity of individual BHs with magnetic dipoles may be estimated as $L_* \sim (2\mu_*^2 \Omega_*^4)/(3c^3)$. This gives
\begin{equation}
\frac{L_*}{L_w} \sim \left(120\frac{r_s}{R_{\rm lc,*}}\right)^2 \hat a^{15} \sim 0.4 \left(\frac{r_s}{R_{\rm lc,*}}\right)^2 \left(\frac{\hat a}{0.5}\right)^{15}.
\end{equation}
One can see that even though $L_* \gg L_w$ when $\hat a \gg 1$, at coalescence ($\hat a < 1$), $L_*$ becomes smaller than $L_w$. In the slow-balding scenario of \cite{lyutikov11}, the field would evolve into a monopole configuration. In this case, one may estimate $L_* \sim (\Omega_*\mu_*/r_s)^2/c \sim (c^5/G) \hat q^2$. This gives 
\begin{equation}
\frac{L_*}{L_w} \sim 2400 \hat a^{15} \sim 0.07 \left(\frac{\hat a}{0.5}\right)^{15}.
\end{equation}
Again thanks to the strong dependence of $L_w$ on $\hat a$, $L_*$ becomes negligibly small compared with $L_w$ at $\hat a < 1$.

\section{Radio and gamma-ray emission}

In this model, radio emission may be produced in the inner magnetosphere through coherent ``bunching'' curvature radiation mechanism by the pairs streaming out from the magnetosphere, similar to the case of radio pulsars. The time scale (Eq.(\ref{eq:tr1})) sets an upper limit on the duration of an FRB. To reproduce a typical FRB luminosity $L_{\rm FRB} \sim 10^{41}~{\rm erg / s}$, the requirement of $L_w > L_{\rm FRB}$ (from Eq.(\ref{eq:Lw1})) gives $\hat q > 3\times 10^{-8}$ for $\hat a=1$ and $\hat q > 2 \times 10^{-10}$ for $\hat a=0.5$.

The magnetic field configuration of the dynamical magnetosphere is complicated. For simplicity, we adopt a dipole field as an order of magnitude estimate. Right before the coalescence, one has $a = (4GM/c^2) \hat a = (1.8 \times 10^7 ~{\rm cm}) (M/30 M_\odot) \hat a$ and $\hat a \geq 1$. For a dipole field line $r = r_e \sin^2\theta$, one may take $r_e \sim a$ right before the coalescence (which implies a nearly isotropic emission beam). Noticing that the curvature radius $\rho \sim (0.3-0.6) r_e$ in a wide range of $r$, one may approximate $\rho \sim 0.45 r_e \sim (8 \times 10^6~{\rm cm}) (M/30 M_\odot) \hat a$. 
The typical curvature radiation frequency of the pairs is
\begin{equation}
\nu = \frac{3}{4\pi} \frac{c}{\rho} \gamma_e^3 \simeq (0.9 \times 10^9 ~{\rm Hz})~\hat a^{-1} \left(\frac{M}{10 M_\odot}\right)^{-1} \gamma_{e,2}^3,
\end{equation}
where the Lorentz factor of the pairs $\gamma_e$ is normalized to 100, the nominal Lorentz factor value of pairs from a pulsar polar cap cascade \citep[e.g.][]{zhangharding00}. This frequency is the typical frequency of the observed FRBs. The curvature radiation emission power of an electron is $P_e = \frac{2}{3} \frac{e^2 c}{\rho^2} \gamma_e^4 \simeq (7.2 \times 10^{-15} ~{\rm erg~s^{-1}})~ \hat a^{-2} (M/10 M_\odot)^{-2} \gamma_{e,2}^4$. For the bunching coherent mechanism \citep{ruderman75b}, the total emission power is $P = N_{\rm bunch} N_e^2 P_e$, where $N_e$ is the number of electrons in each bunch, $N_{\rm bunch}$ is the number of bunches, with the total number of electrons defined by $N_{\rm tot} = N_{\rm bunch} N_e$. The minimum number of electrons that are needed to reproduce the typical luminosity of an FRB, $L_{\rm FRB} = 10^{41}~{\rm erg~s^{-1}}~ L_{\rm FRB,41}$, can be derived by assuming that $N_{\rm bunch} = 1$ and $N_{\rm tot} = N_e$, so that $N_{\rm tot,min}  = ({L_{\rm FRB}}/{P_e})^{1/2}  \simeq  3.7 \times 10^{27}~ \hat a ({M}/{10 M_\odot}) \gamma_{e,2}^{-2} L_{\rm FRB,41}^{1/2}$.
The total number of emitting electrons in the magnetosphere may be estimated as
$N_{\rm tot} \sim  Q/e \simeq (2.1 \times 10^{31}) q_{-9}$,
which is $\gg N_{\rm tot,min}$ even if $\hat q$ is normalized to $10^{-9}$. This suggests that energetically the bunching mechanism is able to power an FRB in such a transient magnetosphere.

The pair cascade process only converts a small fraction of the wind energy into radio emission.
The dominant energy component in the outflow would be in the form of a Poynting flux. The EM energy is entrained in the outflow and would be dissipated at a large radius through magnetic reconnection triggered by internal collision or current instabilities \citep{zhangyan11,lyutikov03}. Assuming that gravitational waves (GWs) travel with the speed of light\footnote{The GW 150914 indeed leads the putative associated GRB by 0.4 s \citep{connaughton16}. This would give the tightest constraint on the Einstein's Equivalent Principle (EEP) to date \citep{wu16}.}, the FRB is essentially simultaneous with the GW chirp signal, but the $\gamma$-ray emission would be slightly delayed with respect to the GW chirp signal due to the slightly smaller speed of the Poynting flux with respect to the speed of light. Suppose that the GRB emission starts at radius $R_1$ with Lorentz factor $\Gamma_1$ and ends at radius $R_2$ with Lorentz factor $\Gamma_2$, one may define 
\begin{equation}
t_1 = \frac{R_1}{2\Gamma_1^2 c}, ~~~ t_2 = \frac{R_2}{2\Gamma_2^2 c}.
\end{equation}
Several observational time scales can be estimated as follows:
\begin{itemize}
\item The delay time between the onset of the GRB and the final GW chirp signal is
\begin{equation}
\Delta t_{\rm GRB} \sim (t_1 - \tau_{1.5}) (1+z).
\label{eq:tGRB}
\end{equation}
\item The rising time scale of the GRB is defined by
\begin{equation}
t_r \sim {\rm max} (\tau_{1.5}, t_2 - t_1) (1+z) .
\end{equation}
\item The decay time scale of the GRB is defined by
\begin{equation}
t_d \sim t_2 (1+z).
\end{equation}
\item The total duration of the GRB is
\begin{equation}
\tau = t_r + t_d.
\end{equation}
\end{itemize}

\section{GW 150914 and the possible associated GRB}

\cite{connaughton16} reported a weak, hard X-ray transient that was potentially associated with GW 150914. The false alarm probability is 0.0022, and the poorly-constrained localization is consistent with that of GW 150914. The putative GRB has a duration $\tau \sim 1$ s, and was delayed with respect to the GW signal by $\Delta t_{\rm GRB} \sim 0.4$ s. Assuming the redshift of GW 150914 \citep{GW150914}, $z = 0.09^{+0.03}_{-0.04}$, the 1 keV - 10 MeV luminosity of the putative GRB is $1.8^{+1.5}_{-1.0} \times 10^{49}~{\rm erg~s^{-1}}$. 

The properties of this putative short GRB may be interpreted by our model. According to Eq.(\ref{eq:Lw1}), one can estimate the required charge of the BHs as
\begin{equation}
 \hat q_{-4} \simeq 3.5 \hat a^{15/2} \eta_\gamma^{-1/2} \simeq 0.02 \left(\frac{\hat a}{0.5}\right)^{15/2} \eta_\gamma^{-1/2},
 \label{eq:q-4}
\end{equation}
where $\eta_\gamma = L_\gamma/ L_w$ is the radiative efficiency of the GRB, which ranges in (0.1-1) for known GRBs \citep{zhang07a}. 
According to Eq.(\ref{eq:muOmega}), the required $\mu_*\Omega_*$ value is of the order of that of a millisecond magnetar if $\hat q \sim 10^{-5}$, achievable for a rapidly spinning BH. So the putative GBM signal associated with GW 150914 could be interpreted with this model. There are suggestions that the GBM signal may not be real \citep[e.g.][]{greiner16,xiong16}. If so, one may place an upper limit on $\hat q$ of the order of $10^{-5}$. The non-detection of $\gamma$-ray signals from LVT 151012 and GW 151226 \citep{racusin16,smartt16} could pose an upper limit on $\hat q$ to the same order.

The delay and the short duration of the GBM transient with respect to GW 150914 could be readily explained. According to Eq.(\ref{eq:tr1}), approximating $M \sim 30 M_\odot$ for both BHs in GW 150914, one may estimate $\tau_{1.5} \lesssim 
5$ ms, which is $\ll$ the delay time scale $\Delta t_{\rm GRB} \sim 0.4$ s. One therefore has $t_{\rm GRB} \sim t_1$ (noticing $(1+z) \sim 1$), which gives a constraint on the onset radius of emission
\begin{equation}
R_1 \sim 2 \Gamma_1^2 c t_{\rm GRB}  = (2.4 \times 10^{14}~{\rm cm})~\left(\frac{\Gamma_1}{100}\right)^2 \left(\frac{\Delta t_{\rm GRB}}{0.4~{\rm s}}\right).
\end{equation}
The weak signal does not allow a precise measurement of $t_r$ and $t_d$. In any case, the pulse is asymmetric \citep{connaughton16} with $t_d = t_2 \gg t_r = t_2 - t_1$, consistent with the theory. The total duration is $\tau = 2 t_2 - t_1 \sim t_2$, which defines the decay time scale due to the angular spreading curvature effect. One can then estimate the radius where emission ceases, i.e. 
\begin{equation}
R_2 \sim 2 \Gamma_2^2 c t_2 \sim 2 \Gamma_2^2 c \tau  = (6.0 \times 10^{14}~{\rm cm})~\left(\frac{\Gamma_2}{100}\right)^2 \left(\frac{\tau}{1~{\rm s}}\right).
\end{equation}
Even though the Lorentz factor $\Gamma$ for such kind of GRBs is unknown, we can see that for nominal values ($\Gamma_1 \sim \Gamma_2 \sim 100$) of known GRBs \citep{liang10}, the emission radius is much greater than the photosphere radius, suggesting that the GRB emission comes from an optically thin region. The large radius is consistent with the expectation of the models that invoke magnetic dissipation in a Poynting flux dominated outflow \citep{zhangyan11,lyutikov03}.

\section{Event rate densities}

For $\hat q = 10^{-9} - 10^{-8}$ needed to produce FRBs, the required BH $\mu_* \Omega_*$ is $\sim (10^{32}-10^{34}) {\rm G~cm^3~s^{-1}}$, which is much smaller than that of a millisecond magnetar. This suggests that a moderately spinning BH with a moderate magnetic field in a merger system could make an FRB. One would expect more associations of BH-BH mergers with FRBs than GRBs.

The inferred event rate density of BH-BH mergers from the detections of GW150914, GW151226 and LVT151012 \citep{GW-rate} is  $\sim (9-240) ~{\rm Gpc}^{-3}~{\rm yr}^{-1}$. The FRB event rate density may be estimated as
\begin{eqnarray}
\dot \rho_{\rm FRB} & = &  \frac{365  \dot N_{\rm FRB}}{(4\pi/3) D_{z}^3} \simeq (5.7 \times 10^3~{\rm Gpc^{-3}~yr^{-1}}) \nonumber \\
& \times &  \left(\frac{D_z}{3.4~{\rm Gpc}}\right)^{-3} \left(\frac{\dot N_{\rm FRB}}{2500}\right),
\end{eqnarray}
where $\dot N_{\rm FRB}$ is the daily all-sky FRB rate which is normalized to 2500 \citep{keane15}, and $D_z$ is the comoving distance of the FRB normalized to 3.4 Gpc ($z=1$). One can see that the FRB rate is at least 20 times higher than the BH-BH merger rate (see also \citealt{callister16}). Recently \cite{keane16} claimed a cosmological origin of FRB 150418. \cite{spitler16}, on the other hand, reported repeating bursts from FRB 121102, which point towards an origin of a young pulsar, probably in nearby galaxies \citep[e.g.][]{cordes16,connor16}. Based on radio survey data, \cite{vedantham16} suggested that the fraction of cosmological FRBs with bright radio afterglow as FRB 150418 should be a small fraction of the entire FRB population. Our analysis suggests that the BH-BH mergers can account for the cosmological FRBs if their fraction is less than 5\%, and if all BH-BH mergers can have $\hat q$ at least $10^{-10}-10^{-8}$. If the radio transient following FRB 150418 \citep{keane16} is indeed the afterglow of the FRB \citep[cf.][]{williams16,lizhang16}, then the observation is consistent with the prediction of this model \citep{zhang16}.

\section{Summary and discussion}

For BH-BH mergers, if at least one of the BHs carries a certain amount of charge, the inspiral process generates a loop circuit, which induces a magnetic dipole. The rapid evolution of the magnetic moment of the system leads to a magnetospheric outflow with an increasing wind power. If $\hat q$ can be as large as $\sim (10^{-9}- 10^{-8})$, the magnetospheric wind right before the coalescence may produce an FRB, and the BH-BH mergers may contribute to some cosmological FRBs. If $\hat q$ could be as large as $\sim (10^{-5}-10^{-4})$, a short-duration GRB may be produced. The putative short GRB signal associated with GW 150914 \citep{connaughton16} may be interpreted with this model.

The near-isotropic nature of the magnetosphere wind conjectured in this model suggests that every BH-BH merger should be accompanied by an EM counterpart (if $\hat q$ is large enough). The detection of an FRB (or even a GRB) associated with future BH-BH merger GW events would verify this model, and lead to a measurement to $\hat q$ (since the luminosity is essentially a function of $\hat q$ only). The non-detections of GRBs and FRBs associated with these mergers, on the other hand, would place an upper limit on $\hat q$ allowed for astrophysical BHs.

The same physical picture naturally applies to NS-NS and NS-BH merger systems as well. Since those systems have at least one NS, it is guaranteed that at least one member of the merger system carries a $\hat q$ large enough to produce cosmological FRBs (see also \cite{wang16} for an alternative trigger mechanism). The detectable event rate of these mergers, however, is not much larger than BH-BH mergers, since in a large solid angle of such a merger, the FRB could not escape due the absorption of the dynamical ejecta launched during the merger. In systems with larger $\hat q$, the pre-merger dynamical magnetospheric activities would make a possible hard electromagnetic transient leading the main episode of the short GRB (see also a recent discussion on this aspect by \cite{metzger16}). A detection or an upper limit on this signal would give interesting constraints on the properties of the pre-merger systems.




\acknowledgments
I thank the referee, Anatoly Spitkovsky, for constructive comments and criticisms, and Mitch Begelman, Zi-Gao Dai, Tong Liu, Peter M\'esz\'aros, Kohta Murase, Martin Rees, Scott Tremaine, Z. Lucas Uhm, Shao-Lin Xiong, and Bin-Bin Zhang for helpful comments and discussion. This work is partially supported by NASA NNX15AK85G and NNX14AF85G.

\end{document}